# Deriving a lattice model for neo-Hookean solids from finite element methods

Teng Zhang[*]

Department of mechanical and aerospace engineering, Syracuse University, Syracuse, NY, 13244, USA

[*]Corresponding author. Email: tzhang48@syr.edu

**Abstract**

Lattice models are popular methods for simulating deformation of solids by discretizing continuum structures into spring networks. Despite the simplicity and efficiency, most lattice models only rigorously converge to continuum models for lattices with regular shapes. Here, we derive a lattice model for neo-Hookean solids directly from finite element methods (FEM). The proposed lattice model can handle complicated geometries and tune the material compressibility without significantly increasing the complexity of the model. Distinct lattices are required for irregular structures, where the lattice spring stiffness can be pre-calculated with the aid of FEM shape functions. Multibody interactions are incorporated to describe the volumetric deformation. We validate the lattice model with benchmark tests using FEM. The simplicity and adoptability of the proposed lattice model open possibilities to develop novel numerical platforms for simulating multiphysics and multiscale problems via integrating it with other modeling techniques.

## 1. Introduction

Lattice models, at continuum levels, are usually referred to numerical methods of simulating deformation of solids with a system a discrete units interacting via springs [1-8], which are also known as spring networks. They have been widely used to various solid materials, ranging from rocks [9] and concretes [10, 11] to composites [12-14]. For soft materials (e.g., elastomers and gels), lattice models have been used to drive continuum constitute laws from microstructures of polymer networks [15-18]. The well-known Arruda-Boyce model is derived using a cubic unit cell with 8 Langevin chains [15]. Recent studies have employed lattice models to describe non-uniform deformation in soft materials [19-22]. For instance, a gel lattice spring method (gLSM) has been developed by linking the spring deformation energies with strain energy in FEM [21, 22]. The gLSM has been successfully utilized to study self-oscillating, multiphysics interactions and shape changes of chemo- and thermo-responsive gels [21-24]. In another study, an irregular tetrahedral nonlinear truss network model was proposed to link microscale chain statistical mechanics to macroscale deformation [20], which can capture the nonlinear stress-strain curves in rubbers [20] and shape memory polymers [25]. This can be seen as a bottom-up fashion of deriving lattice models from molecular chain levels.

The simplicity and efficiency of the lattice model make it very promising to solve various mutlphysics and multiscale coupling problems, where the challenges mainly come from integrating different modeling techniques to capture the interactions across different scales and phases [26-31]. For example, lattice models and lattice Boltzmann method have been coupled to understand the interaction between soft particles and microchannels and patterned surfaces [32-36]. Furthermore, a lattice model can be seen as a generalized coarse-grained (CG) models that seamlessly matches CG models for polymer networks at microscales [37-39] and can also achieve the same accuracy as FEM simulations in describing large deformations macroscales [21, 22]. Therefore, the lattice model provides a very promising means to bridge molecular and FEM simulations for various mechanical challenges that both molecular structure and nonlinear bulk deformation are important and strongly coupled, like the mechanics of interfaces in soft materials [40]. However, most lattice models can only be proved equivalent to FEM for regular lattices, and this may hinder the application of lattice models.

In this paper, we show that an irregular lattice can also be equivalent to FEM for neo-Hookean solids by calculating distinct lattice spring stiffness from FEM shape functions and adopting an averaged volumetric strain in the calculation of bulk deformation energy. The strain energy for the neo-Hookean solids can be computed from the deformation energies associated with length change in lattice springs linking two different nodes in the lattice and area/volume variation of the lattice, which correspond to two body and multibody interaction, respectively. These energy functional can be easily implemented into conventional FEM and molecular dynamical (MD) simulations. This can substantially facilitate the coupling between lattice models with other modeling techniques, such as coarse-grained polymer chain model [37], lattice Boltzmann method [41, 42], and smoothed-particle hydrodynamics [43-45] and helps to leverage the high parallel efficiency of open source software, like LAMMPS [46] and PETSc [47], to run large scale simulations of the mutlphysics and multiscale coupling problems.

## 2. FEM based lattice models

We start our discussion with two-dimensional (2D) problems. The strain energy density ($U$) of a neo-Hookean solid can be written as [48]

$$U = \mu(I_1 - 3)/2 + \lambda(\ln J)^2/2 - \mu \ln J, \tag{1}$$

where $\mu$ is the shear modulus, $\lambda$ is the Lame constant, $I_1 = \lambda_1^2 + \lambda_2^2 + 1$ for plane strain deformation is the first invariant of the right Cauchy-Green deformation tensor, $\lambda_i, i = 1,2$ is the principal stretches and $J$ is the determinant of the deformation gradient tensor **F**.

The key idea of lattice models is to calculate the strain energy density based on chain (lattice spring) stretching and area change in a lattice (Fig. 1). For example, the energies associated with $I_1$ and volumetric deformation $J$ in a square lattice structure are

$$U_{I_1} = \frac{1}{2}\mu(I_1 - 3) = \frac{\mu}{2A_0}[(r_{12}^2 + r_{23}^2 + r_{34}^2 + r_{14}^2)/6 + (r_{13}^2 + r_{24}^2)/3 - 2], \tag{2}$$

$$U_J = \frac{1}{2}\lambda(\ln J)^2 - \mu \ln J = \frac{1}{2}\lambda(\ln(A/A_0))^2 - \mu \ln(A/A_0), \tag{3}$$

where $A$ and $A_0$ represent the areas of the lattice at the deformed and reference configurations, respectively. It has been shown that Eq. (2) is exact equivalent to the strain energy calculated from FEM with a square linear element [21].

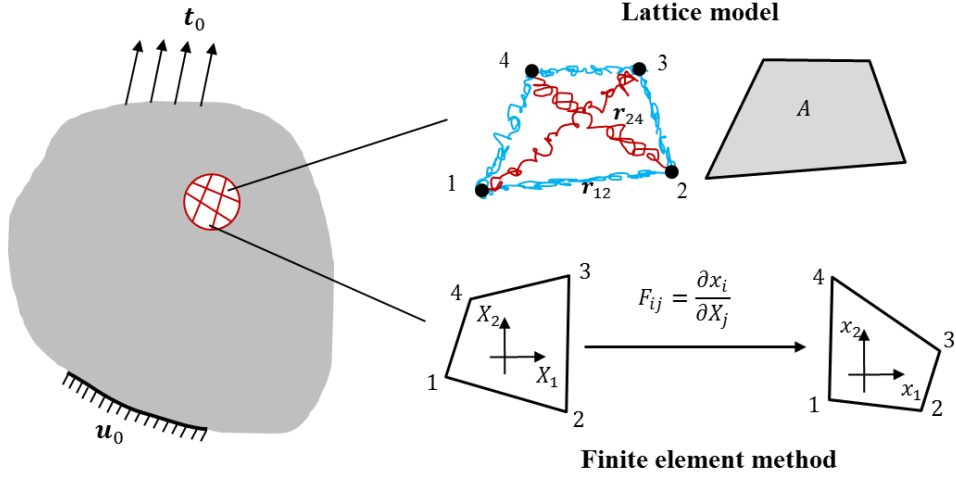

**Fig. 1.** Schematics of the lattice model and finite element method for solving a boundary value problem.

Here, we show that the equivalence between lattice models and FEM can also be established for irregular lattices. To prove that, we first compute the strain energy in the FEM framework, in which the deformation gradient tensor in an irregular element need to be first calculated

$$F_{ij} = \frac{\partial x_i}{\partial X_j} = x_i^a \frac{\partial N^a}{\partial X_j}, \tag{5}$$

where $N^a(X_1, X_2)$ is the shape function, $a = 1,2,3,4$, $i,j,k = 1,2$. Since we have $I_1 = F_{ij}F_{ij} + 1$ for 2D plane strain problems, the strain energy associated with $I_1$ inside the element can be computed as

$$A_0 U_{I_1} = \int \frac{1}{2}\mu \left( x_i^a x_i^b \frac{\partial N^a}{\partial X_j}\frac{\partial N^b}{\partial X_j} - 2 \right) dA_0 = -\frac{1}{2}k_{ab}x_i^a x_i^b - \mu A_0, \tag{6}$$

where $k_{ab} = -\int \mu \frac{\partial N^a}{\partial X_j} \frac{\partial N^b}{\partial X_j} dA_0$. It can be seen from Eq. (6) that $U_{I_1}$ is a quadratic function of nodal positions. Furthermore, we notice that the FEM shape functions are partition of unity (i.e., $\sum_{a=1}^{4} N^a = 1$), and this leads to the following relations

$$\sum_{a=1}^{4} k_{ab} = -\mu \int (\sum_{a=1}^{4} \frac{\partial N^a}{\partial X_j}) \frac{\partial N^b}{\partial X_j} dA_0 = 0, b = 1,2,3,4, j = 1,2. \qquad (7)$$

Substituting Eq. (7) into Eq. (6) and using the symmetry of $k_{ab}$, the strain energy can be rewritten as summation of energies in springs linking different nodes in the lattice structure,

$$U_{I_1} = \frac{1}{2} A_0^{-1} \sum_{b=2, b>a}^{4} \sum_{a=1}^{3} k_{ab} r_{ab}^2, \qquad (8)$$

where $r_{ab} = x^a - x^b, a = 1,2,3, b = 2,3,4$ represent lattice springs between node $a$ and $b$, and $k_{ab}$ is the corresponding spring stiffness. For irregular lattices, $k_{ab}$ can be pre-calculated through Gaussian quadrature with isoparametric mapping. The calculation of volumetric energy in Eq. (3) is the same as the averaged area strain used in the F-bar method [49]. Since the model contains both two body (lattice springs) and multibody interaction (area penalty), it can be seen as a generalized lattice model.

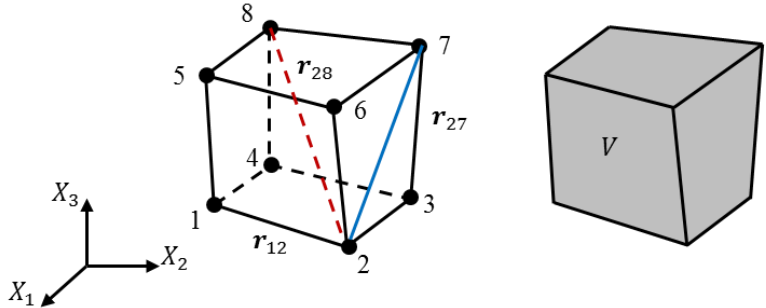

**Fig.2.** The three-dimensional lattice model.

A three-dimensional (3D) hexahedron lattice is a straightforward generalization of the 2D case discussed above. The strain energy associated with $U_{I_1}$ of a 3D hexahedron element can be partitioned into 28 lattice springs in a general case (Fig. 2),

$$U_{I_1} = \frac{1}{2}V_0^{-1} \sum_{b=2,b>a}^{8} \sum_{a=1}^{7} k_{ab} r_{ab}^2 , \tag{9}$$

where $r_{ab} = x^a - x^b$, and $k_{ab} = -\int \mu \frac{\partial N^a}{\partial X_j} \frac{\partial N^b}{\partial X_j} dV_0$, $j = 1,2,...,8, a = 1,2,..7, b = 2,3,...,8$.

Similar to the 2D case, the spring constants $k_{ab}$ can be pre-calculated through Gaussian quadrature. Following the same F-bar concept, the volumetric energy is calculated with an averaged volumetric strain (Fig. 2),

$$U_J = \frac{1}{2}\lambda(\ln(V/V_0))^2 - \mu \ln(V/V_0), \tag{10}$$

where $V$ and $V_0$ represent the volumes of the lattice at the deformed and reference configurations, respectively. The details of creating initial input information and performing simulations in lattice models will be discussed in Sec. 3.

It should be pointed out that Balazas and her collaborators have laid out foundations of connecting FEM and lattice models [21, 22]. By further modeling pressure as an independent variable, they have successfully applied the lattice model with regular lattices (both in 2D and 3D) to solving various nonlinear and multiphysics problems in incompressible gels [21-24]. Here, we further show that the lattice model can be established for irregular lattice if different lattice spring constants can be assigned based on the shape of the lattice. The volumetric energy is now calculated in the fashion of volume average, making it possible to capture a wide range materials from compressible to nearly incompressible.

## 3. Numerical examples

We have established the theoretical framework of simulating deformation of a neo-Hookean solid with the lattice model, which may contain irregular lattices. In this section, we will validate the proposed lattice model by comparing several benchmark tests with FEM, including tension of a constrained elastic layer, cook membrane test, and bending of a 3D beam with a triangle cross-section. In these examples, we adopt the following unit system: mm for length, mN for force, and kPa for stress. In addition, we assume the shear modulus in the simulations are all equal to 1kPa.

Before we apply the lattice model to complicated problems, we first check the effective spring stiffness for a rectangular lattice structure with a unit length and height of $h$, as shown in Fig. 3(a). The corresponding values of the spring stiffness for each of the six springs are

$$k_{12} = k_{34} = \mu\left(\frac{1}{3}h - \frac{1}{6h}\right), k_{23} = k_{34} = \mu\left(\frac{1}{6}h - \frac{1}{3h}\right), k_{13} = k_{24} = \mu\left(\frac{1}{6}h + \frac{1}{6h}\right). \quad (11)$$

There are totally three different types of springs due to the symmetry of the structure, whose spring stiffness is denoted as $k_{12}$, $k_{23}$, and $k_{24}$, respectively. It clearly shows that the values of spring stiffness depend on the shape of the lattice. It can be verified that the values of stiffness in Eq. (11) are equal to those in Eq. (2) if we set $h=1$. For more general $h$, these values of spring stiffness are shown in Fig. 3(b). One interesting feature is that the effective spring stiffness can be negative (i.e., $k_{12} = -\frac{1}{6}\mu$ for $h = \frac{1}{2}$), although the total strain energy associated with $I_1$ ($U_{I_1}$) is the summation of these six springs and positive definite. This indicates that the lattice springs in the current model may not be necessarily related to physical springs.

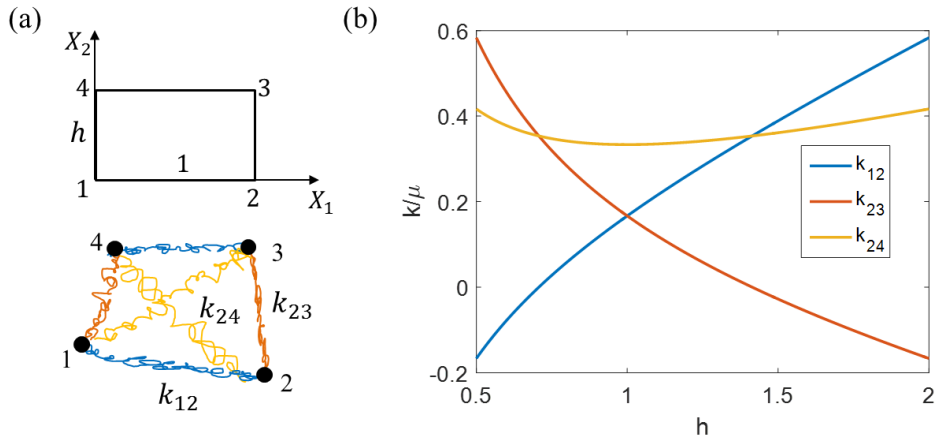

**Fig. 3.** The lattice model with a rectangular shape. (a) The initial geometry and deformed lattice structure. (b) The values of three spring stiffness in the lattice model as a function of height $h$.

For each lattice simulation, we utilize the information of nodal positions and element connectivity in a FEM model to compute the values of spring stiffness inside each element based on Eq. (7). Therefore the lattice model and FEM can share the same geometry information, such that a lattice structure in the lattice model can be associated with an element in FEM. It should be noted that these values of spring stiffness are only dependent on initial nodal positions inside each lattice structure and do not require further Gaussian quadrature for simulating deformations of the structure. After creating the input structures in lattice models, residual forces and tangential stiffness matrix can be formed by assembling contribution from stretching of lattice springs and area change of lattices, in a similar way to FEM.

The first numerical example is to simulate a constrained elastic layer under tension, as shown in Fig. 4a. The elastic layer is under 2D plane strain deformation, which has rectangular shape with initial length $H_0 = 10$ and length $L_0 = 2H_0$. The elastic layer is discretized into small rectangular lattices, whose heights are half of the lengths. The elastic layer is fixed at the bottom edge and stretched by applying displacement condition at the top edge (Fig. 4(a)). We also perform FEM simulation with ABAQUS [50] to predict the deformed configuration of the elastic layer and reaction force at the top edge. A user defined material subroutine is implemented in ABAQUS to incorporate the strain energy formula defined in Eq. (1). The ABAQUS simulations are using 4-node bilinear element with reduced integration, hourglass control, and hybrid with constant pressure (CPE4RH).

We discretize the elastic layer into 400 elements in FEM, and will correspondingly have 400 lattices in the lattice model. An overlaid configuration from lattice model and FEM is shown in Fig. 4(a), indicating excellent agreement between these two methods under large deformation (overall stretch ratio is 2). The reaction forces at the top edge from lattice model and FEM are compared in Fig. 4(b). It can be seen that the lattice model can reproduce results from FEM for materials with different compressibility, including $\lambda = 5\mu, 10\mu$, and $100\mu$, which corresponds to Poisson's ratio as 0.417, 0.455, and 0.495.

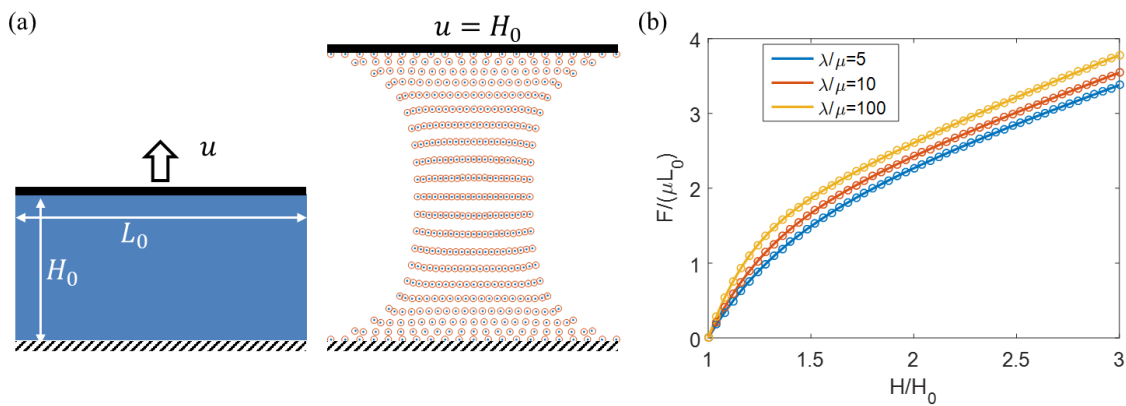

**Fig. 4.** (a) The initial configuration of the elastic layer and the deformed configurations from lattice model (dot) and FEM (circle). The simulations have 20 element/lattice per side. (b) The reaction forces from lattice model and FEM as functions of stretch ratios and Poisson's ratios.

We next apply the lattice model to the standard cook membrane test for volumetric locking. To describe the nearly incompressible materials, we set $\lambda = 100\mu$, corresponding to a Poisson's ratio of 0.495. The left side is fixed, and a uniformly distributed traction $\boldsymbol{t}_0 = [0\ 0.25]^T$ is applied at the right side. The traction is defined in the reference configuration. We start the simulations with 4 elements/lattices per side and uniformly refine the mesh up to 64 element/lattices per side. The deformed configurations from both methods are overlaid again in Fig. 5 (a), where each side has 16 elements/lattices. Except visualizing the comparison between the two methods, we also plot the vertical displacement of the right top corner (C) as we increase the number of elements/lattices. It can be seen that lattice model can converge to the reference solution and produce very close results compared to FEM. We do see some differences especially for coarse meshes, and think it can be attributed to the hybrid element used in ABAQUS.

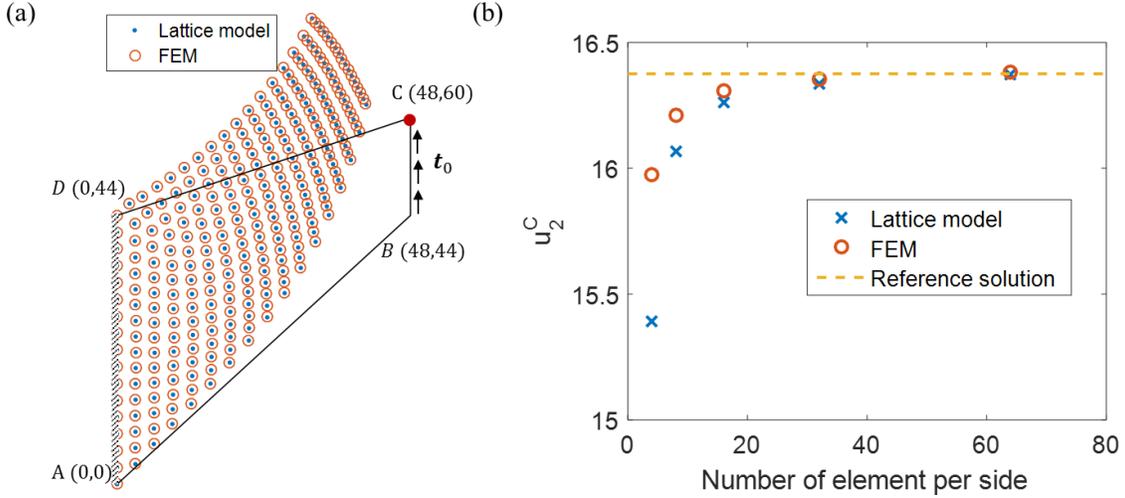

**Fig. 5.** (a) The initial configuration of the cook membrane (solid line) and the deformed configurations from lattice model (dot) and FEM (circle). (b) Vertical displacements at node C from lattice model and FEM. The reference solution is obtained from running high order FEM (CPE8RH) using the finest mesh (64x64).

Our last example is a 3D beam structure with a triangle cross section with length $L_0 = 200$ and edge width $b = 11.57$, as shown in Fig. 6(a). The left end of the 3D beam is fixed, and a concentrated force $\boldsymbol{F} = [0, f, 0]^T$ is applied to node $A$. A nearly incompressible neo-Hookean is adopted by setting $\lambda = 100\mu$. We further normalized all the forces and stresses by the shear modulus $\mu$. For the FEM simulation, we employ the 8-node linear brick element with reduced integration, hourglass control, and hybrid with constant pressure (C3D8RH). The average mesh size is 2, and this generates a total 2700 elements/lattices in the simulations. The deformed configuration from the lattice model is shown in Fig. 6 (b) at $f = 0.05$. The color indicate the value of vertical displacement. We compare the vertical displacement of point $A$ in lattice model and FEM (Fig. 6(c)) and find very good agreement with each other.

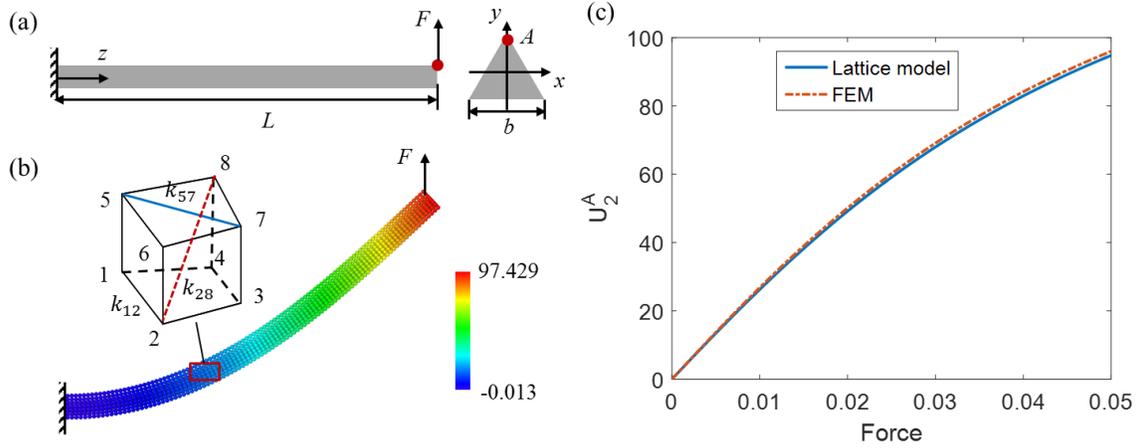

**Fig. 6.** (a) Schematics of the 3D beam with triangle cross section. (6) The deformed configuration from lattice model with a representative 3D lattice. The color indicates vertical displacement. (c) Vertical displacements at point *A* from lattice model and FEM.

## 5. Conclusion

We show that lattice models can be derived from FEM for neo-Hookean solids defined by a widely used strain energy function. The equivalence is rooted in the quadratic nature of the first invariant of the right Cauchy-Green deformation tensor and the partition of unity property of shape functions in FEM. In addition, a wide range Poisson's ratio can be achieved by adopting an averaged volumetric strain in the calculation of bulk deformation energy. The proposed lattice model retains the simplicity of previous model for regular lattice structures and can handle complicated geometries. We validate the lattice model by comparing benchmark tests for large deformations in 2D and 3D neo-Hookean solids with ABAQUS simulations. The intent here is not to promote the lattice model as a competitor of FEM, as the lattice model itself does not provide new functions beyond FEM. Instead, our studies provide a rethinking of the classical lattice models and extend their capabilities of handling different geometries and describing a wide range of material compressibility. The attraction of the lattice model is its potential to be integrated with other modeling techniques to tackle challenging problems with multiscale and multiphysics coupling and interactions.

**Acknowledgement**

The simulations were performed at the Triton Shared Computing Cluster (TSCC) at SDSC and Comet cluster at SDSC in XSEDE (No. TGMSS170004) supported by NSF.